\documentstyle [aps,twocolumn,epsf]{revtex}
\catcode`\@=11 
\newcommand{\be}{\begin{equation}}
\newcommand{\ee}{\end{equation}}
\newcommand{\bea}{\begin{eqnarray}}
\newcommand{\eea}{\end{eqnarray}}

\def\ket#1{\left\vert #1 \right\rangle}

\def\matel#1#2#3{\langle #1 \vert #2 \vert #3 \rangle}
\def\inprod#1#2{\langle #1 \vert #2 \rangle}

\def\etal{{\it et al.\/}}

\begin{document}
\title{Spin-Gap Physics, Ground State Degeneracy, and Bound States \\ on
the Depleted Kagom{\'e} Lattice}

\author{C.~Hooley, A.~M.~Tsvelik}

\vspace{0.5cm}

\address{Department of Physics, University of Oxford, 1 Keble Road, Oxford OX1 3NP, UK}

\vspace{3cm}

\address{\rm (Received: )}
\address{\mbox{ }}
\address{\parbox{14cm}{\rm \mbox{ }\mbox{ }
We analyse the antiferromagnetic spin-${1 \over 2}$
Heisenberg model on a depleted kagom{\'e}
lattice, where some bonds have been reduced to
exchange integral $J_2 \ll J_1$.  The fully
depleted system consists of
1D chains, each with a doubly degenerate
singlet-pair ground state and a spectral gap (like
the Majumdar-Ghosh model).  There are localised and itinerant low-energy excitations.
The modes from the lowest branch of excitations are incapable of lifting the 2D system's ground
state degeneracy at finite $J_2/J_1 \ll 1$.
Low-energy excitations of the 2D system are dominated by
coherently propagating bound states of the 1D
excitations.
}}
\address{\mbox{ }}
\address{\parbox{14cm}{\rm PACS No: 75.10.Jm}}
\maketitle

\makeatletter
\global\@specialpagefalse
\makeatother
{\bf INTRODUCTION.}

In recent years, interest in systems exhibiting a spin gap in the
excitation spectrum has been fuelled by the appearance of spin-gap
phenomena in various situations, including frustrated spin systems\cite{numerics}.
The analysis of toy models of spin-gap physics is
noticeably easier in one spatial
dimension, so considerable attention has been devoted to this line of work.

Despite the well known
distinction between integer and half-integer spin
excitation spectra, some spin-${1 \over 2}$ systems with intermediate
interaction ranges have a spin gap:\ the Majumdar-Ghosh (MG)
model\cite{mg1,mg2} and several
related models\cite{oth1,oth2} exhibit a rich variety of behaviour.
The study of the low-lying excitations of the MG
model has proved a non-trivial task\cite{exc1,exc2,exc3,exc4,exc5}, and we need a
spin-gapped model analogous to the MG one, but in which
the low-lying excitations are easily tractable.  We
present such a model, obtained by
removing certain bonds from the spin-${1 \over 2}$ Heisenberg model
on a kagom{\'e} lattice (cf.\ the approach of \cite{mila} to the full
2D kagom{\'e} antiferromagnet).

Under this depletion, the lattice decouples into a set of
one-dimensional chains, which have exact singlet-pair ground states
(like the MG model) and whose low-lying excitation
spectrum can be obtained analytically.  The ground state of each chain
is doubly degenerate, and consequently the ground state of the whole
system has an degeneracy exponential in the number of chains.

We then reinstate
the removed bonds at a strength $J_2 \ll J_1$, where $J_1$ is the
on-chain exchange integral.  This introduces a branch of domain wall--domain wall bound states
which propagate coherently on the 2D lattice, and have
lower energy than the on-chain modes.
Furthermore, we show that the ground state degeneracy obtained for
$J_2/J_1=0$ {\sl is not lifted\/} by the modes in the lowest-lying band for finite $J_2/J_1 \ll 1$.

\vspace{4mm}
{\bf THE MODEL.}

\begin{figure}
\begin{center}
\leavevmode
\hbox{\epsfxsize=6.5cm \epsffile{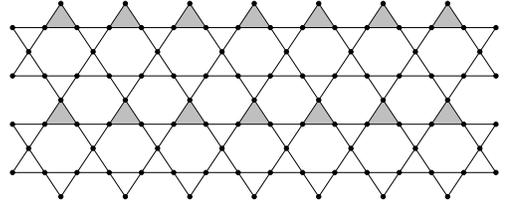}}
\end{center}
\caption{The depleted kagom{\'e} lattice.  The grey triangles of bonds
have been weakened to exchange integral $J_2$; the remainder have exchange
integral $J_1 \gg J_2$.}
\label{depleted}
\end{figure}

The Hamiltonian of our model is given by
\be
H = J_1 \! \! \sum_{\langle i,j \rangle \not\in A} \! \! {\bf S}_i \cdot {\bf S}_j
\, + \, J_2 \! \! \sum_{\langle i,j \rangle \in A} \! \! {\bf S}_i \cdot {\bf S}_j,
\label{ham}
\ee
where the ${\bf S}_i$ are vectors of spin-${1 \over 2}$ operators.
It represents the system shown in Fig.~\ref{depleted}, where the set
$A$ contains the bonds on the edges of the grey triangles.  These are
the bonds that have been weakened to strength $J_2/J_1$ with respect
to the remainder.
{\bf THE ONE-DIMENSIONAL CHAINS.}

\begin{figure}
\begin{center}
\leavevmode
\hbox{\epsfxsize=6.5cm \epsffile{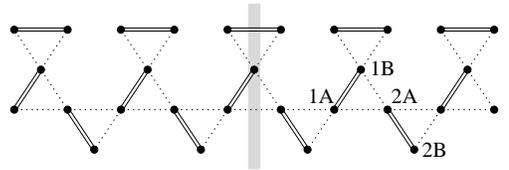}}
\end{center}
\caption{The red ground state of the chain.  Reflection
about the grey bar yields the degenerate blue ground state.}
\label{gs}
\end{figure}

{\bf Ground States.}  In the limit $J_2/J_1 = 0$, the 2D system decouples into a
set of 1D chains.
Each chain has doubly degenerate ground states
made up of nearest-neighbour singlet pairs.  One ground state (called the `red' ground state for convenience) is shown in Fig.~\ref{gs}.  The
other (`blue') ground state is obtained by a mirror
reflection of the red state about the grey bar in the figure.
These are clearly eigenstates of the on-chain part of the Hamiltonian
(\ref{ham}), since each bond is either occupied by a singlet, or part
of a triangle with a singlet on one side; see the work of Klein\cite{Klein} and
Affleck \etal\cite{Affleck} on the relationship between total-spin projection operators and
valence bond crystal states.  In the former case, an elementary calculation gives
$$
{\bf S}_1 \cdot {\bf S}_2 \left [ {1 \over \sqrt{2}} \left(
\ket{\uparrow \downarrow} - \ket{\downarrow \uparrow} \right) \right] =
- {3J_1 \over 4} \left [ {1 \over \sqrt{2}} \left(
\ket{\uparrow \downarrow} - \ket{\downarrow \uparrow} \right) \right],
$$
whereas the contributions of the other two sides of the triangle
cancel out:
$$
\left( {\bf S}_1 \cdot {\bf S}_3 + {\bf S}_2 \cdot {\bf S}_3 \right)
\left[ {1 \over \sqrt{2}} \left( \ket{\uparrow \downarrow X} -
\ket{\downarrow \uparrow X} \right) \right] = 0,
$$
where $X {=} \uparrow, \downarrow$.
Notice that the ground states of Fig.~\ref{gs} are orthogonal to each other only in
the limit when the number of spins on the chain tends to infinity,
since $\inprod{\,{\rm blue}\,}{\,{\rm red}\,} \sim 2^{-N}$,
where $3N$ is the number of spins.

The bonds at the top of the chain are occupied by singlet
pairs in both ground states.  If we ignore them for the time being
(see below), we are left with a model rather like a
Majumdar-Ghosh chain, except that only alternate sites have
next-nearest-neighbour interactions.

This model, obtained by removing the top bonds from our depleted kagom{\'e} chain,
is called the ``delta chain'', or ``sawtooth lattice'', and has been analysed by several
authors.  Early work\cite{Hamada,Doucot,Kubo} was numerical,
and established the existence of
an energy gap; later studies\cite{Nakamura,Sen} concentrated on analytic work
(mainly variational calculations),
modelling the low-lying excitations as domain walls of a few spins.
We summarise the domain-wall physics of a single chain here, in a notation suitable to the
remainder of the letter.

{\bf Low-Energy Excitations.}  It is easy to show that a localised excitation introduced into a red
ground state will not propagate coherently.  To do this, we consider
a group of four spins on the chain (1A, 1B, 2A, and 2B --- see Fig.~\ref{gs}), and rewrite the local
Hamiltonian in a basis consisting of two-spin eigenstates.  Namely,
\begin{eqnarray}
\ket{S}_n & \equiv & {1 \over \sqrt{2}} \left( \ket{\uparrow \downarrow}
- \ket{\downarrow \uparrow} \right), \quad \ket{-}_n \equiv \ket{\downarrow \downarrow}, \nonumber \\
\ket{0}_n & \equiv & {1 \over \sqrt{2}} \left( \ket{\uparrow \downarrow}
+ \ket{\downarrow \uparrow} \right), \quad \ket{+}_n \equiv \ket{\uparrow \uparrow},
\end{eqnarray}
where the subscript $n$ refers to the bond between the spins
described.  In this case, we consider bonds 1 and 2.

\begin{figure}
\begin{center}
\leavevmode
\hbox{\epsfxsize=8cm \epsffile{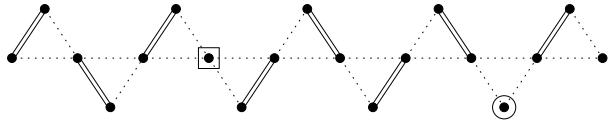}}
\end{center}
\caption{The two types of domain wall on the chain:\ stationary
(boxed) and mobile (circled).}
\label{excitons}
\end{figure}

Initially, the state of these bonds will be given by
$\ket{S}_1 \ket{S}_2$.
Henceforth we shall drop the subscripts 1 and 2, leaving them implicit
in the order of the kets.
Now let us consider introducing a local excitation by modifying the
state of bond 1.  It is easy to show that
\begin{eqnarray}
\hat{H} \left( \ket{+} \ket{S} \right) & = & {J_1 \over 4} \ket{+} 
\ket{S} + {J_1
\over 2} \ket{+} \ket{0} - {J_1 \over 2} \ket{0} \ket{+}, \nonumber \\
\hat{H} \left( \ket{0} \ket{S} \right) & = & {J_1
\over 4} \ket{0} \ket{S} + {J_1 \over 2} \ket{+} 
\ket{-} - {J_1 \over 2} \ket{-} \ket{+}, \nonumber \\
\hat{H} \left( \ket{-} \ket{S} \right) & = & {J_1
\over 4} \ket{-} \ket{S} + {J_1 \over 2} \ket{0} 
\ket{-} - {J_1 \over 2} \ket{-} \ket{0},
\end{eqnarray}
while $\ket{S} \ket{S}$ is an eigenstate of
$\hat{H}$.  Since there is no term on the right hand side
of the form $\ket{S} \ket{X}$, coherent
propagation of such a localised excitation cannot occur.
These considerations lead us to conclude that the low-energy excitations of
this chain are of the domain-wall type.  If we begin with an initially
red chain, we may introduce a region of blue ground state into it,
bounded by two domain walls of distinct natures (see
Fig.~\ref{excitons}).

The stationary domain wall (boxed in the figure) is
flanked by singlets, and hence remains spatially localised.  The mobile domain wall
(circled) feels the effect of the on-chain Hamiltonian, and so acquires a dispersion.
We take our basis states to be $\ket{j}$, representing a mobile
domain wall at site $j$ on the chain.
The calculation of the dispersion relation is modified by the fact
that these basis states are not mutually orthogonal.  Consequently, an
expansion of our eigenstate in these basis states,
$\ket{\psi} = \sum_j c_j \ket{j}$,
yields an equation for the vector $c_j$ which reads ${\bf H} {\bf c} = E {\bf A} {\bf c}$,
where $H_{ij} = \matel{i}{H_{\rm chains}}{j}$ and $A_{ij} =
\inprod{i}{j}$.  The energy spectrum is therefore obtained from the
eigenvalues of the effective Hamiltonian ${\bf A}^{-1} {\bf H}$.

Hence we find that the dispersion relation
for the mobile wall is given by
\be
E(k) = {7J_1 \over 6} \left( {5 \over 4} + \cos k \right),
\label{onespec}
\ee
where the ground state is taken to have zero energy.  This result is independent of
the correlation (singlet or triplet) between the mobile and stationary
domain walls, so this branch contains magnetic and non-magnetic modes.

{\bf Local Modes.}  As well as these domain wall modes, the spectrum
contains modes in which the spin pairs on the top horizontal bonds are
excited into a triplet state, whose r{\^o}le must be considered.
However, compared to the gap of the domain wall band ($\approx J_1/4$), the
singlet-triplet gap on these bonds ($=J_1$) is rather large, so we propose
to ignore excitations of the top bond singlets when considering low-energy processes
on the chain.

\vspace{4mm}
{\bf EXCITATIONS OF THE 2D MODEL.}

{\bf Pair-Hopping Amplitude.}  Analysing the effect of $\alpha \not = 0$
on the excitations of the one-dimensional
chains leads naturally to the consideration of tunnelling
processes; however, it is clear that a single mobile wall cannot tunnel by
itself, as this would violate the restriction that the chain
states be unchanged at large distance.

However, there exists an amplitude for pair hopping of a
stationary and a mobile domain wall.  To describe it, let us introduce
a set of basis states more general than those used in the 1D analysis.  In our new notation, $\ket{i,j}_1$ represents a state
of chain 1 with a stationary domain wall at site $i$, and a mobile
domain wall at site $j$.
Our previous basis states $\ket{j}$ are now represented by
$\ket{0,j}_1$.

\begin{figure}
\begin{center}
\leavevmode
\hbox{\epsfxsize=8cm \epsffile{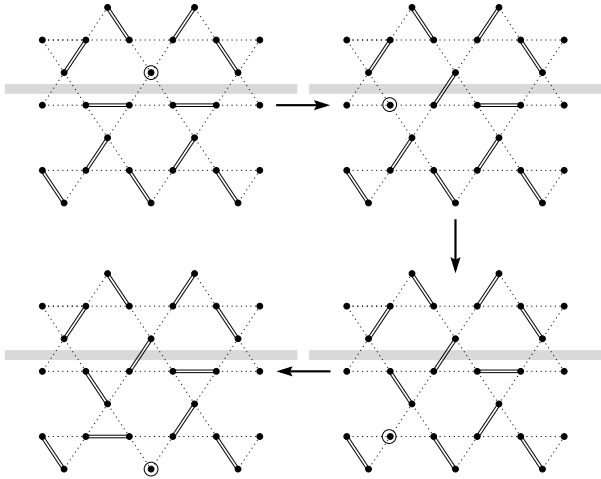}}
\end{center}
\caption{A pictorial representation of the steps in the inter-chain
tunnelling process.  Those triangles through which the grey bar passes
have bonds with exchange integral $J_2$.}
\label{tunnel}
\end{figure}

Let us deal with the case where both chains are in the red
ground state at large distance.
There is a term of perturbation theory, first order in $J_2/J_1$, that
connects the states
$\ket{i,j}_1 \ket{\,{\rm red}\,}_2$ and
$\ket{\,{\rm red}\,}_1 \ket{k,j}_2$.
This term is represented diagrammatically in Fig.~\ref{tunnel}.
Using the standard formulae of Rayleigh-Schr{\"o}dinger perturbation
theory, and approximating the energy of the local mode by $J_1$, we may
obtain the amplitude
\be
A_{\rm nonrefl} = {J_2 \over 20} \left( -2 \right)^{-\vert i-j \vert-\vert k-j
\vert} \Theta\left((i-j)(k-j)\right),
\label{nramp}
\ee
where $\Theta(x)$ is the unit step function.  The precise value of the
prefactor depends on the value we use for the local mode gap, but the
linearity in $J_2$ is a basic feature.

This amplitude corresponds to the hopping of a mobile domain wall from site $j$
on chain 1 to the same site on chain 2, while the stationary wall hops
from site $i$ on chain 1 to a (potentially different) site $k$ on
chain 2.  The step function guarantees that the mobile wall is on the
same side of the stationary wall after the hopping as it was before.
This treatment applies equally to the case where the two ground states
are blue.  Both of these cases will be called `non-reflective
hopping', since the sign of $k$ is preserved in the hopping process.
The alternative situation, `reflective hopping', will be taken up in a
forthcoming publication\cite{prb}; we do not expect our qualitative
results to be altered by the introduction of reflective hopping (see
concluding discussion).

{\bf Bound States (Non-Reflective Hopping).}
To analyse the non-reflective case, we formulate an effective model by
introducing creation and annihilation operators for the mobile
($c^\dag_{n,i},c_{n,i}$) and stationary ($d^\dag_{n,i},d_{n,i}$)
domain walls; here, $n$ and $i$ are chain and site indices
respectively.  It is clear that the domain walls must obey an exclusion principle, but
this still leaves a choice of statistics:\ fermions or hard-core bosons?  In the dilute limit,
the exchange statistics are not relevant to
the calculation; conventionally, we choose the operators $\left\lbrace
c,c^\dag,d,d^\dag \right\rbrace$ to obey Fermi statistics.

In these terms, the effective Hamiltonian is
\bea
{\hat H}_{\rm eff} & = & t \sum_{n,i} \left( c^\dag_{n,i} c_{n,i+1} +
{\rm h.c.} \right) \nonumber \\
& & + \, (\varepsilon-\mu) \sum_{n,i} c^\dag_{n,i} c_{n,i} + \, \mu \sum_{n,i} d^\dag_{n,i} d_{n,i} \nonumber \\
& & + \sum_{n,i,j,k} \left( V^{n-1,n}_{ijk} c^\dag_{n,i} d^\dag_{n,k}
d_{n-1,j} c_{n-1,i} + {\rm h.c.} \right)
\label{effmod}
\eea
where $t={7 \over 12}J_1$, $\varepsilon={35 \over 24}J_1$,
and $V_{ijk}$ is just the amplitude $A_{\rm nonrefl}$ from
Eq.~(\ref{nramp}), with $i$ being the position of the mobile wall, $j$
that of the stationary wall before hopping, and $k$ that of the
stationary wall after hopping.

Because of the translational invariance in the $x$ (parallel to the
chains) and $y$ (transverse to the chains) directions, we may Fourier
transform and work in momentum space.  The action is given by
\bea
S & = & \sum_{\omega,p,h} c^\dag_{\omega ph} \left( \omega - \varepsilon - 2t\cos
p + \mu \right) c_{\omega ph} \nonumber \\
& & + \sum_{\omega,p,h} d^\dag_{\omega ph} \left( \omega - \mu
\right) d_{\omega ph} \nonumber \\
& & + \sum_{\def\arraystretch{0.5} \begin{array}{c} \scriptstyle \omega_1 \omega_2 \omega_3 \\
\scriptstyle pqr \\
\scriptstyle h \end{array}} \Gamma^{(0)} {\def\arraystretch{0.5} \begin{array}{l}
\scriptstyle h \\ \scriptstyle pqr \\
\scriptstyle \omega_1 \omega_2 \omega_3 \end{array}} c^\dag_{\omega_3rh}
d^\dag_{\omega_4sh} d_{\omega_2qh} c_{\omega_1ph},
\eea
where $\omega_4 \equiv \omega_1 + \omega_2 - \omega_3$, $s \equiv
p+q-r$, and $c^\dag_{\omega ph}$, $c_{\omega ph}$, $d^\dag_{\omega ph}$,
$d_{\omega ph}$ now represent Grassmann fields.  The $\omega_i$ are fermionic Matsubara frequencies,
$p,q,r,s$ are momenta along the chains, and $h$ is the transverse
momentum.  Recall that, in the case we consider, all momenta are
restricted to positive values.  The bare vertex part, $\Gamma^{(0)}$, is given by
\bea
\Gamma^{(0)} & = & {J_2 \over 20} {\cos h \over 2\pi} \left( f(q)f(-s)
+ f(-q)f(s) \right),
\eea
where the function $f(q)$ is the Fourier transform of the hopping amplitude's
spatial dependence, given in Eq.~(\ref{nramp}).

The formation of bound states is signalled by the appearance of
divergences in the dressed vertex part, $\Gamma$.  Being unable to
solve the full set of Schwinger-Dyson equations, we make a simple Ansatz:\ we assume the self-energies of the
two propagators to be zero.  Our defence is that the gap in
the unperturbed model is quite large.

The remaining bound state condition can be solved analytically only near two lines in the
Brillouin zone, $h=\pm {\pi \over 2}$.  Setting $h={\pi \over
2}-\delta$, we find that
\be
\epsilon = - {J_2 \delta \over 10t} \left( 2\cos^2 P - 4\cos P + 3
\over \left( 4\cos P - 5 \right)^2 \right),
\label{height}
\ee
where $P$ is the on-chain momentum, and ${1 \over 2} \epsilon^2$ is the distance that the bound state
lies {\sl below\/} the band of unperturbed modes, i.e.
$E_{\rm bound} = {7 \over 24} J_1 - {1 \over 2} \epsilon^2$.
Similar behaviour is found near the line $h=-{\pi \over 2}$.  We
tentatively conclude that the branch of bound states exists throughout
the Brillouin zone, touching the bottom of the continuum band
tangentially along the lines $h=\pm {\pi \over 2}$.

\vspace{4mm}
{\bf GROUND STATES OF THE 2D MODEL.}

In the limit $J_2/J_1=0$, the double degeneracy of
each chain in the system leads to a total degeneracy $g = 2^L$,
where $L$ is the number of chains.

It is noteworthy that we have been unable to find a process (within the manifold of domain wall
states) that lifts this ground state degeneracy in the $\alpha \not = 0$ case.
Any such process consists of a finite number of inter-chain hopping events.
Each such hop between two chains of the same ground state has a counterpart hop between two chains of
different ground states, the only difference being that in the latter case the left-to-right order of
the stationary and mobile domain walls is reversed.  Due to a combination of the spatially extended nature
of the domain wall states (see above) and the imposition of periodic
boundary conditions, such a reversal cannot make a difference to the amplitude of the process.  Hence all
perturbation theory corrections to the energy of a given ground state of the two-dimensional lattice have
counterparts of equal magnitude and sign for any other ground state.

Although no explicit calculation has been performed, the degeneracy is
almost certainly lifted by perturbative
processes involving the excitation of local modes on the `top'
horizontal bonds of the chains.
Nonetheless, it is interesting to note that, because of the effective screening of
the chains from one another
due to the singlets on the top bonds, the modes in our low-energy manifold are
incapable of lifting the
ground state degeneracy of the $\alpha = 0$ case.

\vspace{4mm}
{\bf CONCLUSION.}

We have presented a quasi-1D chain with nearest-neighbour
antiferromagnetic exchange and a spin-gapped excitation spectrum that
can be treated analytically in the low-energy limit.  This chain was obtained by a natural depletion
of the 2D kagom{\'e} antiferromagnet, and as well as
itinerant and localised domain walls (obtained in \cite{Nakamura,Sen} by similar
methods to ours) it has local modes with a gap larger than that of the
itinerant excitations.

Recoupling these chains weakly, we have obtained a pair-hopping
amplitude of the itinerant and localised 1D modes that arises
naturally from the nature of the 1D ground states.  Furthermore, we
have shown that the ground state degeneracy of the 2D system (which is
exponential in the number of chains) is not lifted by processes in the lowest branch
for finite
$J_2/J_1 \ll 1$; this effect is due to symmetry properties of the ground states and
the nature of the interchain coupling.  As regards the excitations,
the pair hopping amplitude leads to a band of bound states of the 1D
domain walls:\ these bound states propagate coherently on the 2D
lattice, and have lower energy than the on-chain excitations.

While our model is not proposed as a starting point for treating the
isotropic 2D kagom{\'e} antiferromagnet, it has several novel and
interesting features, notably the inability of the domain wall modes to lift
the ground state degeneracy\cite{footnote} and the formation of a band of coherently
propagating 2D states for arbitrarily small $J_2/J_1 \not = 0$.

Finally, one might ask whether the consideration of reflective rather
than non-reflective hopping will change our results.  Qualitatively,
one may argue that it will not:\ since bound states already form in the
non-reflective system, the existence of reflective hopping will
act only to strengthen the tendency to bind, perhaps
modifying the prefactor in (\ref{height}).  These
topics, along with a semiclassical method of treating the
non-reflective case, will be taken up in a forthcoming publication \cite{prb}.

\vspace{4mm}
{\bf ACKNOWLEDGMENTS.}

We are pleased to acknowledge useful discussions with
J.\ T.\ Chalker, C.\ Lhuillier, P.\ Lecheminant, P.\ Azaria, R. Siddharthan,
and A.\ Campbell-Smith. 
One of us (CH) acknowledges financial support from EPSRC (UK) studentship
96305551.
 

\vspace{-5mm}
\begin{thebibliography}{99}
\vspace{-16mm}
\bibitem{numerics} C. Waldtmann, H.-U. Everts, B. Bernu,
P. Sindzingre, C. Lhuillier, P. Lecheminant, and L. Pierre,
Euro. Phys. Journal B {\bf 2}, 501 (1998).
\bibitem{mg1} C. K. Majumdar and D. K. Ghosh, J. Math. Phys. {\bf 10}, 1388 (1969).
\bibitem{mg2} C. K. Majumdar and D. K. Ghosh, J. Math. Phys. {\bf 10}, 1399 (1969).
\bibitem{oth1} J. T. Chayes, L. Chayes, and S. A. Kivelson, Commun. Math. Phys. {\bf 123},
53 (1989).
\bibitem{oth2} K. Takano, J. Phys. A (Math. and Gen.) {\bf 27}, L269 (1994).
\bibitem{exc1} B. S. Shastry and B. Sutherland, Phys. Rev. Lett. {\bf 47}, 964 (1981).
\bibitem{exc2} W. J. Caspers, K. M. Emmett, and W. Magnus, J. Phys. A (Math. and Gen.)
{\bf 17}, 2687 (1984).
\bibitem{exc3} K. Takano, J. Phys. Soc. Japan {\bf 63}, 4565 (1994).
\bibitem{exc4} T. Nakamura, S. Takada, K. Okamoto, and N. Kurosawa, J. Phys.:\ Cond. Mat.
{\bf 9}, 6401 (1997).
\bibitem{exc5} H. Yokoyama and Y. Saiga, J. Phys. Soc. Japan {\bf 66}, 3617 (1997).
\bibitem{mila} F. Mila, Phys. Rev. Lett. {\bf 81}, 2356 (1998).
\bibitem{Klein} D. J. Klein, J. Phys. A (Math. and Gen.) {\bf 15}, 661 (1982).
\bibitem{Affleck} I. Affleck, T. Kennedy, E. H. Lieb, and H. Tasaki,
Phys. Rev. Lett. {\bf 59}, 799 (1987).
\bibitem{Hamada} T. Hamada, J. Kane, S. Nakagawa, and Y. Natsume, J. Phys. Soc. Japan
{\bf 57}, 1891 (1988).
\bibitem{Doucot} B. Dou{\c c}ot and I. Kanter, Phys. Rev. B {\bf 39}, 12399 (1989).
\bibitem{Kubo} K. Kubo, Phys. Rev. B {\bf 48}, 10552 (1993).
\bibitem{Nakamura} T. Nakamura and K. Kubo, Phys. Rev. B {\bf 53}, 6393 (1996).
\bibitem{Sen} D. Sen, B. S. Shastry,
R. E. Walstedt, and R. Cava, Phys. Rev. B {\bf 53}, 6401 (1996).
\bibitem{prb} C. Hooley and A. M. Tsvelik, in preparation.
\bibitem{footnote} This is essentially an effect of the
self-generated screening between neighbouring
quasi-1D chains.
\end{thebibliography}
\end{document}